\documentclass[12pt,a4paper]{article}
\usepackage{epsfig}
\begin{document}

\centerline{\Large EUROPEAN ORGANIZATION FOR NUCLEAR RESEARCH}
\vskip 6cm
\Large
\centerline{\bf A novel concept for the detection} 
\centerline{\bf of tau neutrino appearance}
\normalsize
\vskip 2.5cm
\centerline{Roger Forty\footnote{E-mail address:
    Roger.Forty@cern.ch}}
\vskip 4mm
\centerline{\it CERN, Geneva}
\vskip 2.5cm

\begin{abstract}
A novel concept for the detection of tau neutrinos is presented,
potentially suitable for use in a long-baseline neutrino oscillation
experiment. It relies on the direct identification of the tau leptons
produced in charged-current interactions, by imaging the Cherenkov light
that the tau generates in $\rm C_6F_{14}$ liquid. In a simple simulation
about half of the tau leptons can be successfully identified in this way.
\end{abstract}

\newpage\mbox{~}\newpage

\setcounter{page}{1}
\pagestyle{plain}

\section{Introduction}

Strong evidence for neutrino oscillation has come from Super-Kamiokande,
from the zenith angle dependence of the muon deficit that they observe for
atmospheric neutrinos~\cite{SK}.  The favoured explanation is the oscillation
of muon neutrinos to tau neutrinos, which escape detection in their
apparatus.  These results require confirmation using an artificially
generated neutrino beam, and accelerator-based experiments that will check
the muon neutrino disappearance are underway~\cite{K2K,MINOS}.  A
long-baseline beam from CERN to Gran Sasso is also under
discussion~\cite{CNGS}.  As the question of $\nu_\mu$ disappearance should
have been settled by the time that experiments using that beam take data,
the key issue for them is to confirm that tau neutrinos
are indeed being produced. Since the neutral current interactions of the
$\nu_\tau$ cannot be distinguished from those of other neutrinos, the
crucial point is to identify charged-current interactions, $\nu_\tau
N\rightarrow \tau^- X$, by the appearance of the tau lepton.

Two experiments are being proposed to make this measurement:
OPERA~\cite{OPERA} and ICANOE~\cite{ICANOE}.  OPERA seeks to identify the
taus by their characteristic short lifetime, corresponding to an average
decay length of about 1\,mm for the CNGS beam energy spectrum.  They
propose to use an emulsion target to recognise the kink from the tau decay,
building on the expertise accumulated by the CHORUS~\cite{CHORUS} and
DONUT~\cite{DONUT} experiments. ICANOE, on the other hand, intends to
recognise the charged-current $\nu_\tau$ events through kinematical
criteria, involving the missing energy from the neutrinos accompanying the
tau decay, and isolation criteria for the tau decay products, following the
approach pioneered by NOMAD~\cite{NOMAD}. These experiments will be
challenging, as the number of charged-current $\nu_\tau$ interactions
expected in each year of operation of the CNGS beam is only about 30 per
kiloton of sensitive detector mass~\cite{CNGS}, for the oscillation
parameters preferred by the Super-Kamiokande data: $\Delta m^2 = 3.5\times
10^{-3}\,{\rm eV}^2$, $\sin^2(2\theta)=1$~\cite{SKlimit}. Maintaining high
efficiency is therefore crucial, whilst background must be suppressed such
that just a few observed events would correspond to an unambiguous signal.

The detection technique presented in this note is different---the idea is
to directly identify the tau by imaging the Cherenkov light that it
produces.  Cherenkov detectors have already been used in this field:
Super-Kamiokande itself relies on the generation of Cherenkov light in
water, but without focussing. A ring-imaging water Cherenkov detector,
AQUA-RICH~\cite{AQUA}, was originally proposed for Gran Sasso, but
insufficient sensitive mass could fit in the experimental halls, so it is
now being pursued as an atmospheric neutrino experiment sited elsewhere
(with a megaton mass!). Due to the chromatic dispersion in water, a
relatively narrow energy bandwidth is assumed for photon detection in
AQUA-RICH.  Coupled with the 20\% detector coverage this leads to typically
0.5 detected photons per mm of track length, insufficient to see the tau
track. The concept presented here is to use $\rm C_6F_{14}$ liquid as the
radiator, which due to its low dispersion allows a wider photon energy
bandwidth, and to have full detector coverage.  This leads to 13 detected
photons per mm, and direct detection of the Cherenkov ring from the tau
then becomes feasible.  Furthermore, the increased density of the radiator
would allow a kiloton detector to fit comfortably in a Gran Sasso hall.

\section{Detector concept}

\begin{figure}[t]
\begin{center}\epsfig{file=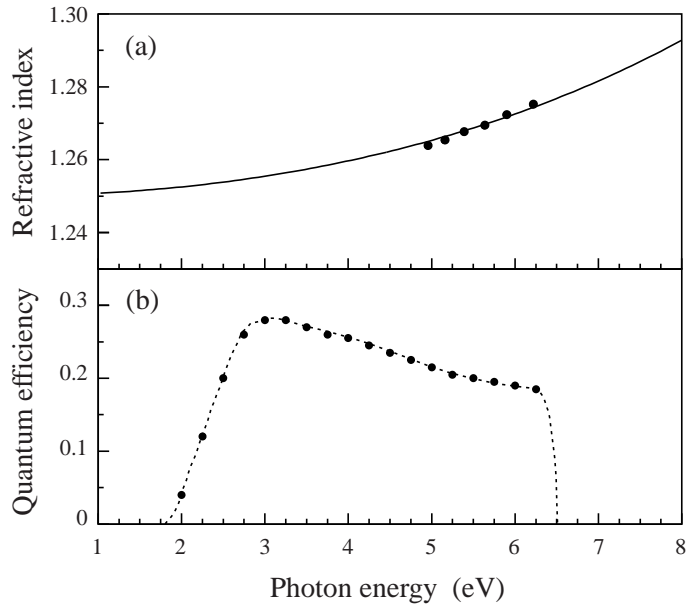,height=8cm}\end{center}\vspace*{-3mm}
\caption{(a) Refractive index versus photon energy for $\rm C_6F_{14}$
  liquid: the points are data taken from~\cite{Jacques}, with a
  superimposed single-pole Sellmeier form, adjusted to agree with unpublished
  data in the visible.  (b)~Assumed quantum efficiency versus photon energy
  for the photodetector: the points are taken from a Hamamatsu
  specification for a bialkali photocathode.
\label{refract}}\end{figure}

Perfluorohexane ($\rm C_6F_{14}$) is a well-established radiator material
for RICH detectors~\cite{DELPHI,CRID}, liquid at room temperature. About a
ton of it is used in DELPHI~\cite{Joram}. It has the nice features for this
application of a refractive index of about 1.27, slightly lower than that
of water, whilst being significantly more dense (1.68\,$\rm g/cm^3$). It is
also, after purification, highly transparent to photons with wavelength
down to 200\,nm and beyond~\cite{Joram}, and has low chromatic dispersion: the
dependence of the refractive index on photon energy is shown in
Fig.~\ref{refract}\,(a).

A classical focussed RICH geometry is adopted, with a spherical mirror
following the radiator and a spherical detection surface sited at radius
\begin{equation}
  r_{\rm d} = \frac{r_{\rm m}}{2}\,
  \frac{\sqrt{1+\frac{9}{16}\sin^2\theta_{\rm
  c}}+\frac{3}{8}\sin^2\theta_{\rm c}}{1-\frac{3}{16}\sin^2\theta_{\rm c}}\ ,
\end{equation}
where $r_{\rm m}$ is the mirror radius of curvature and $\theta_{\rm c}$ is
the Cherenkov angle~\cite{Tom}. The saturated Cherenkov angle in $\rm
C_6F_{14}$ is about $38^\circ$, and so $r_{\rm d} = 0.67\,r_{\rm m}$.

\begin{figure}[t]
\begin{center}\epsfig{file=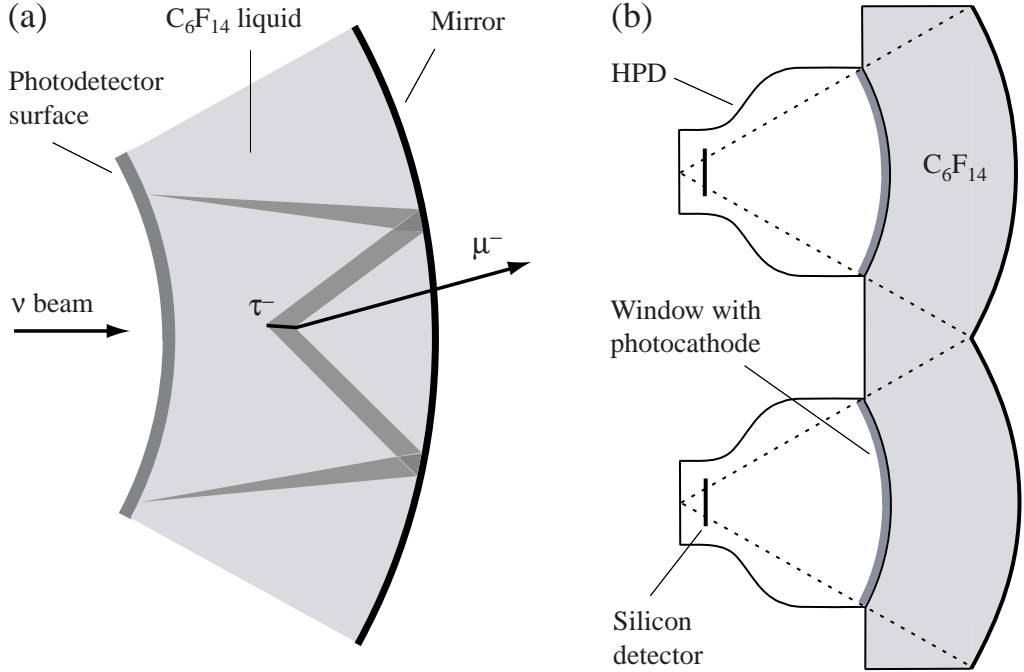,height=9cm}\end{center}
\caption{(a) Schematic layout of a detector module; the focussing of the
  Cherenkov light emitted by the tau is indicated. (b)~A possible
  implementation using large HPDs as the photodetectors.
\label{schematic}}\end{figure}

The tau leptons produced by charged-current interaction of neutrinos from
the CNGS beam are produced in the predominantly forward direction, along
the direction of the beam.  The detector elements are therefore oriented to
collect the light produced by such tracks, as shown schematically in
Fig.~\ref{schematic}\,(a).  The assumed quantum efficiency $Q(E)$ of the
photodetectors is shown in Fig.~\ref{refract}\,(b); it is cut off above
6.2\,eV, corresponding to a quartz entrance window for the detectors.  The
possible implementation of this concept illustrated in
Fig.~\ref{schematic}\,(b) will be discussed in the following section.  For
the purposes of the simulation presented here, a circular detector surface
of 1\,m diameter is assumed, equal to its radius of curvature. The radiator
thickness is then 50\,cm, with a spherical mirror of radius 150\,cm. Such a
module would contain about 0.67\,$\rm m^3$ of $\rm C_6F_{14}$ liquid,
corresponding to about 1100\,kg.  A kiloton detector would thus require
about 900 such modules.

Cherenkov photons produced by the tau and other charged particles in the
event are focussed by the mirror into rings on the detector surface.  For
full detector coverage, the number of detected photoelectrons per ring is
given by:
\begin{equation}
  N = \left(\frac{\alpha}{\hbar c}\right)\, L\,\int
  Q\,T\,R\,\sin^2\theta_{\rm c}\, dE\ ,
\end{equation}
where the factor in parentheses is a constant with value $370\,{\rm
eV^{-1}cm^{-1}}$, $L$ is the track length in the radiator, $T$ is the
transmittance of the radiator, and $R$ is the reflectivity of the mirror
(assumed to be 95\%)~\cite{Tom}. The absorbtion length of purified $\rm
C_6F_{14}$ has been measured to be greater than 100\,cm for
$E<6.2$\,eV~\cite{absorb}, so $T=1$ is assumed here. 
Then Eq.~2 corresponds to 13 detected
photoelectrons per mm of track length, which renders the tau track
visible. The muon from $\tau^- \rightarrow \mu^-\nu_\tau\overline{\nu}_\mu$
decays will typically pass through 25\,cm of radiator, giving 3200
photoelectrons.  The RICH optics result in the position of the rings on
the detector surface being determined by the angle of the tracks,
insensitive to their production point in the radiator, and is thus well
adapted to identification of the tau decay kink.

\begin{figure}[t]
\begin{center}\epsfig{file=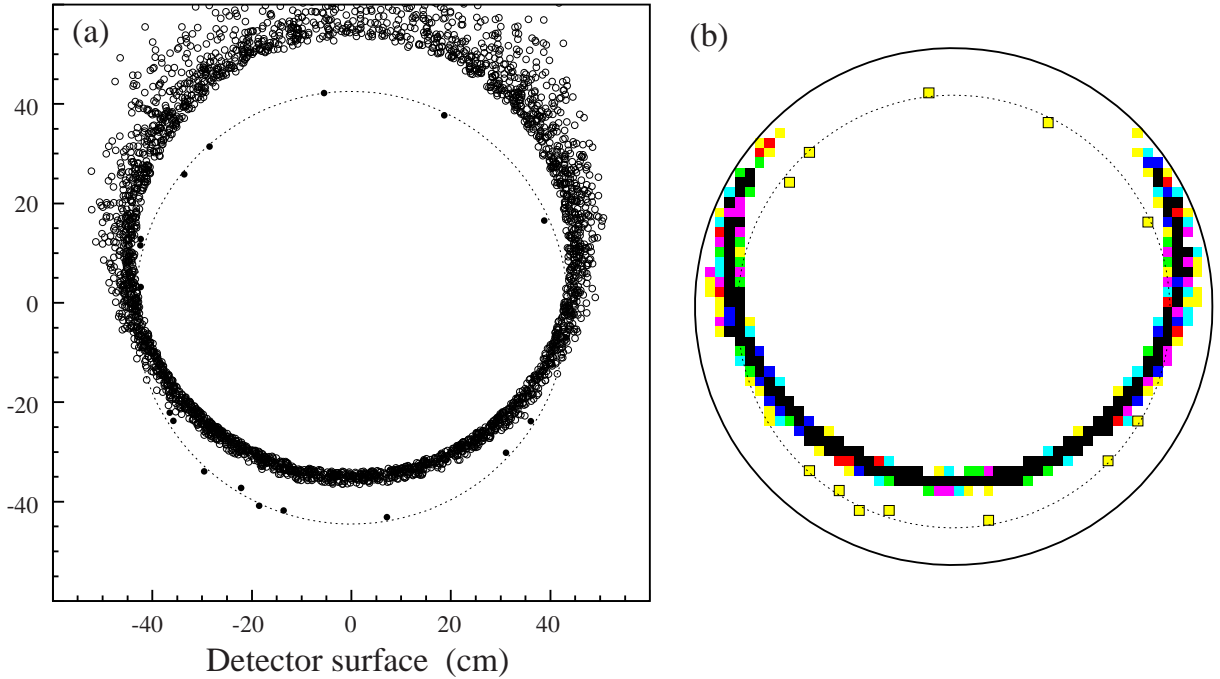,height=9cm}\end{center}\vspace*{-3mm}
\caption{Display of a single $\tau^- \rightarrow
  \mu^-\nu_\tau\overline{\nu}_\mu$ event in the detector module: (a)~impact
  points of the detected photoelectrons on the detector surface; those from
  the tau are marked with solid points, with the ring image indicated by a
  dashed line. (b)~The same event after pixellization of the detector
  surface (within a radius of 50\,cm); the density of shading increases
  with increasing number of detected photoelectrons.
\label{event}}\end{figure}

Such events have been simulated, taking the tau and muon track parameters
from a detailed simulation of quasielastic interactions of a neutrino beam
with the CNGS energy spectrum~\cite{private}.  In quasielastic events $\rm
\nu_\tau n\rightarrow \tau^- p$ the only other track is a proton, which is
below threshold for producing Cherenkov light if it has momentum less than
1.2\,GeV: this is the case for 85\% of the simulated events. For this
simple simulation, multiple scattering of the tracks was ignored, and
photons generated along the track length in the radiator according to the
distribution shown in Fig.~\ref{refract}\,(b), calculating their
Cherenkov angle according to the dispersion curve in
Fig.~\ref{refract}\,(a).  A typical event is shown in
Fig.~\ref{event}\,(a), where the tau decay length was 1.5\,mm and the kink
between tau and muon was 100\,mrad. The signature of such decays will thus
be a densely populated ring from the muon, accompanied by an offset
low-intensity ring from the tau. In the case of $\tau^- \rightarrow \rm
e^-\nu_\tau\overline{\nu}_e$ decays, the electron will shower in the $\rm
C_6F_{14}$, giving a more diffuse ring than for the muon, making the
separation of the tau hits more difficult.  For the single-prong hadronic decay
(corresponding to about half of the tau decays) there is a high probability
of the hadron escaping without nuclear interaction: for the radiator
interaction length of about 55\,cm, a hadronic track of 25\,cm length has
63\% probability of not interacting; these events may therefore also be
useful.

\begin{figure}[t]
\begin{center}\epsfig{file=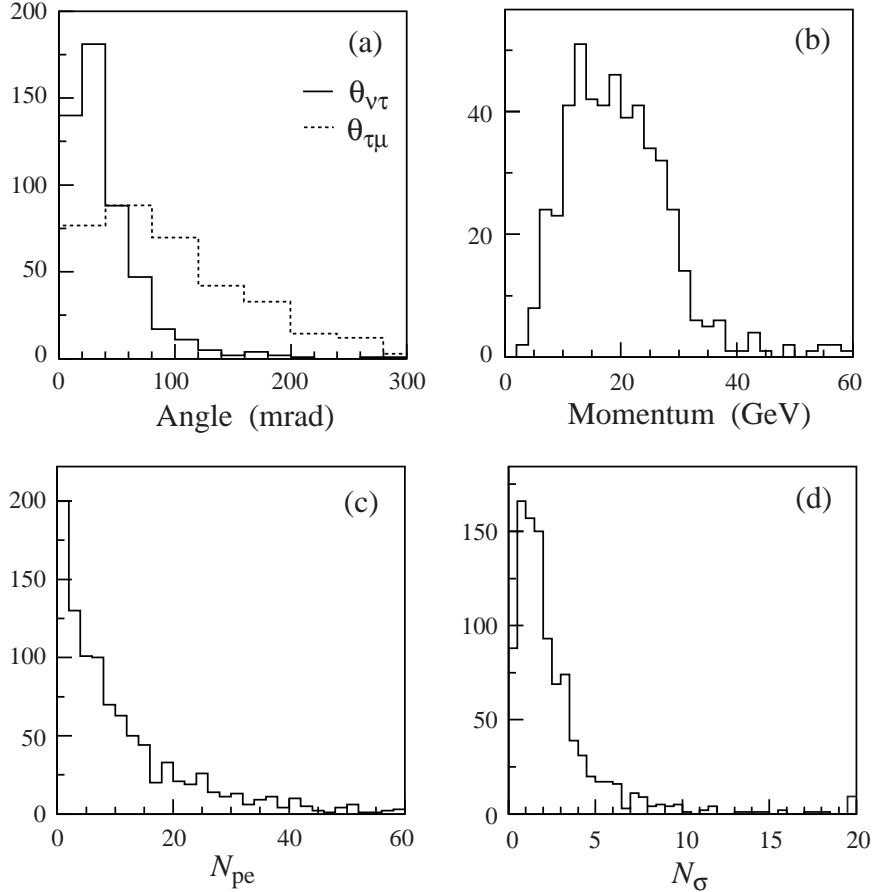,height=12cm}\end{center}\vspace*{-3mm}
\caption{Distributions from simulated tau neutrino interactions:
  (a)~angular distributions of the tau (solid) and muon (dashed), (b)~tau
  momentum, (c)~number of detected photoelectrons from the tau,
  (d)~significance of the difference between the tau hypothesis and the
  saturated Cherenkov angle, for the tau track.
\label{plots}}\end{figure}

The angular distribution of the tau tracks relative to the incoming
neutrino direction in the simulated events is shown as the solid line in
Fig.~\ref{plots}\,(a); as can be seen it is strongly peaked in the forward
direction, with a mean of only 40\,mrad, whilst the distribution of the
kink angle between the tau and muon tracks is broader (dashed in the
figure) with a mean of 150\,mrad.  The momentum distribution of the taus is
shown in Fig.~\ref{plots}\,(b): all are above threshold for generating
Cherenkov light, which is 2.3\,GeV for the tau in $\rm C_6F_{14}$.
The distribution of the number of
detected photoelectrons from the tau track is shown in
Fig.~\ref{plots}\,(c). About 60\% of the events have 6 or more detected
hits, which should be sufficient to recognize the ring.  However, the
selection of tau events can use not only the signature of the offset
low-intensity ring, but also the measurement of the average Cherenkov angle
of the associated photons. For this, good resolution is required to
positively identify the tau by separating its ring radius from that
expected for a saturated track (as would be the case for lighter
particles: e, $\mu$ or $\pi$), as well as to distinguish the photons from
different tracks. A tau with the typical momentum of 20\,GeV emits
Cherenkov light at an angle which is 5\,mrad less than a fully relativistic
particle. 

The Cherenkov angle resolution due to dispersion in the radiator has an RMS
of 10\,mrad per photoelectron, corresponding to 0.7\,cm on the detector
surface.  A detector granularity of $\rm 2\times 2\,cm^2$ is therefore
suitable, to avoid limiting the resolution.  Because of its short decay
length, there is no smearing due to emission-point uncertainty for photons
from the tau. For longer tracks, such as the muon in Fig~\ref{event}\,(a),
there is noticeable effect from spherical aberration, leading to a tail of
photons on the outside of the ring; this, however, tends to be on the side
away from the tau ring, and so should not degrade the pattern
recognition. The result of pixellization of the detector plane is shown for
the same event in Fig.~\ref{event}\,(b).

The determination of the production point of the tau can be achieved by
localizing the muon track as it leaves the detector module, using a
tracking detector. The Cherenkov ring of the muon gives a very precise
measurement of its angle, and the integrated number of photoelectrons
detected on the ring is proportional to the path length.  The tau
production point should be localized in this way to a precision in space of
order 1\,cm, more than adequate for the Cherenkov angle calculation. 

The average Cherenkov angle is determined for all photons from the tau in
each event, and compared to the value expected for a saturated ring. The
significance of the separation, expressed as the number of sigma $N_\sigma$
between the tau hypothesis and the saturated Cherenkov angle, is shown in
Fig.~\ref{plots}\,(d).  Of course, the tau momentum is not fully
reconstructed; nevertheless, the measured muon momentum will provide a
lower limit on the tau momentum, and for the typical muon momenta observed
all light particle types would give a saturated ring.  About half of the
tau tracks have significant separation ($N_\sigma>2$), with 30\% having
$N_\sigma>3$.  The significance could be increased by improving the
resolution with a narrower bandwidth of photon energies, at the cost of
reducing the total number of photoelectrons observed; the optimal cut will
depend on the level of background that needs to be rejected.  The
performance will also be reduced somewhat due to confusion with overlapping
rings from other tracks in the event. In particular, for the muon decays, a
kink angle greater than about 40\,mrad will be required to separate the tau
and muon images; this occurs in about 80\% of events.  Detailed study of
the loss due to pattern recognition awaits a more complete simulation of
the events.

\section{Possible implementation}

\begin{figure}[t]
\begin{center}\epsfig{file=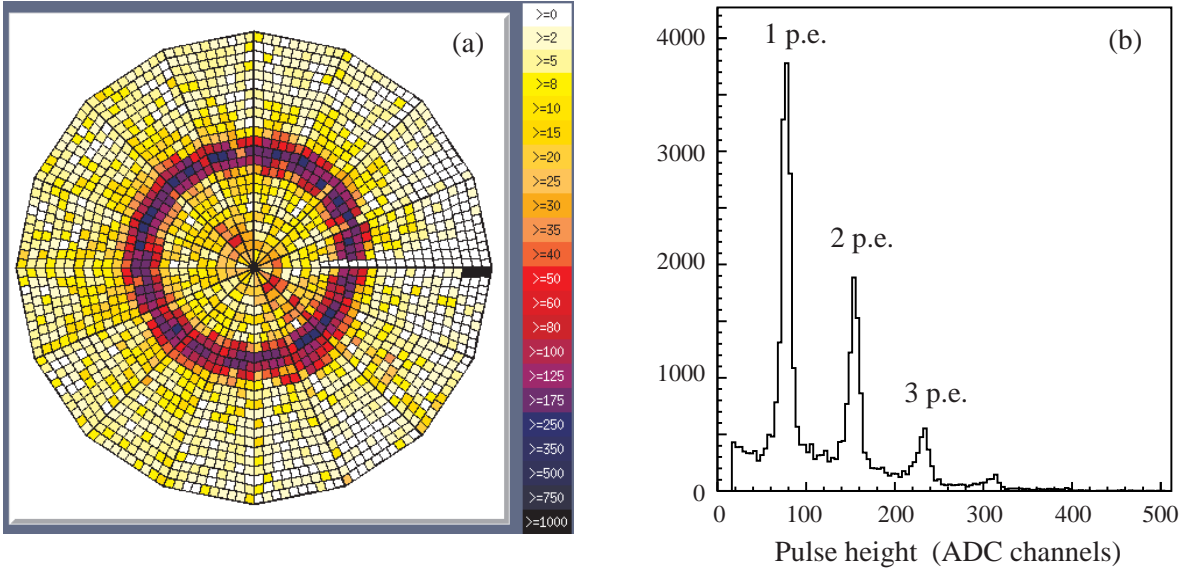,height=8.3cm}\end{center}\vspace*{-4mm}
\caption{(a) Display of beam-test data in the 2048-pad HPD, showing the
  accumulated ring image produced by a pion beam passing through a gas
  radiator; (b)~pulse-height spectrum from an HPD, showing the
  well-separated photoelectron peaks.
\label{HPD}}\end{figure}

To keep high detection efficiency it is advantageous to cover the detection
surface of a module with a single detector.  Since the mass of radiator
that is imaged by a detector scales as the cube of the detector diameter,
the largest possible detectors are desirable to limit the number required.
A 1\,m diameter hybrid photodiode (HPD) detector has been proposed for
AQUA-RICH~\cite{AQUA}. These devices combine the photocathode and focussing
of vacuum photodetectors with the spatial and energy resolution of silicon
detectors. They have been the subject of an intense program of R\&D for the
RICH detectors of the LHCb experiment~\cite{LHCb, pixel}, and one of the
devices developed has 2048 channels in a 5$^{\prime\prime}$ diameter
(127\,mm) envelope~\cite{Pad}.  These tubes are fabricated at CERN with a
bialkali ($\rm K_2CsSb$) photocathode.\footnote{Note that Hamamatsu is
advertising an HPD with GaAsP photocathode that achieves 45\% quantum
efficiency at 500\,nm: such performance could double the number of detected
photoelectrons from the tau.}  A recent test-beam image from one of them is
shown in Fig.~\ref{HPD}\,(a).  Envelopes for a 10$^{\prime\prime}$ version
of this tube have recently been manufactured, and a 20$^{\prime\prime}$
version is already foreseen (the photomultipliers used by Super-Kamiokande
are also of 20$^{\prime\prime}$ diameter). Extrapolation to a
40$^{\prime\prime}$ (1\,m) diameter tube appears feasible.  With 2048
channels, the effective pixel size at the photocathode would be $\rm
2\times 2\,cm^2$, ideal for the present application.  The excellent energy
resolution makes photon counting straightforward in these tubes, as
illustrated in Fig.~\ref{HPD}\,(b).

\begin{figure}[t]
\begin{center}\epsfig{file=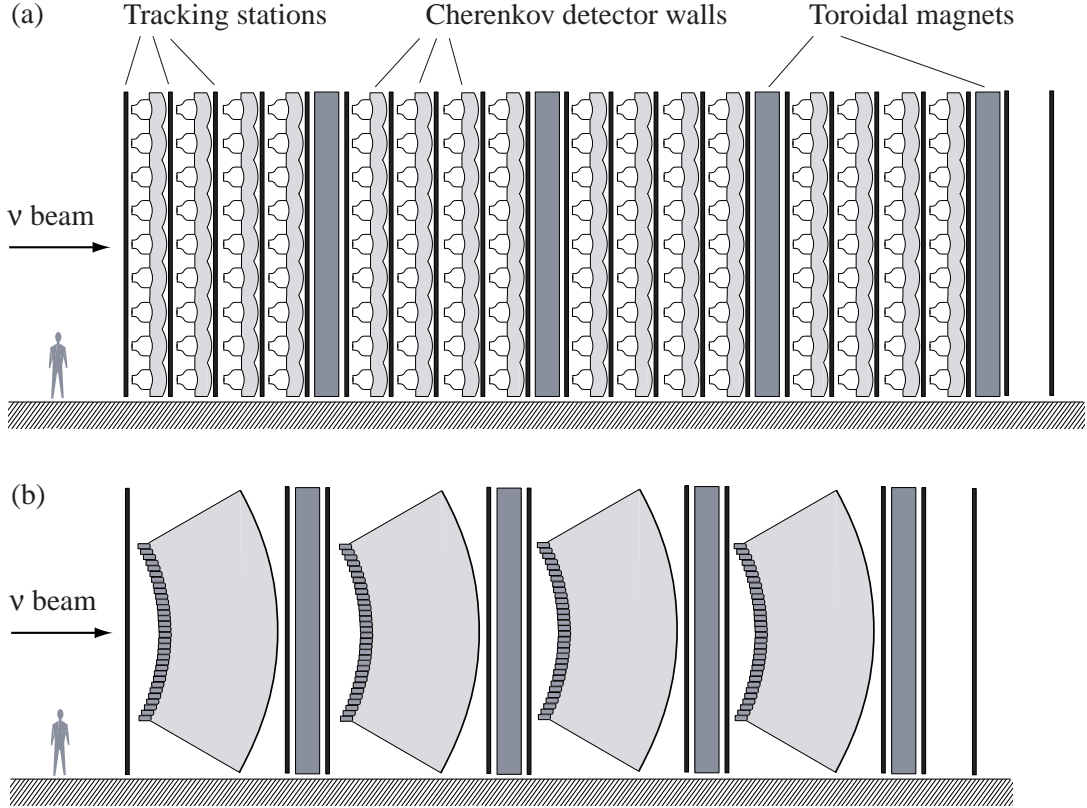,height=11cm}\end{center}\vspace*{-4mm}
\caption{Two possible layouts of detector modules to form a kiloton experiment:
  (a)~using large HPDs for the photodetector, (b)~covering the detection
  surface with many close-packed tubes.
\label{layout}}\end{figure}

With such an HPD as the photodetector, the layout of a module would be as
shown in Fig.~\ref{schematic}\,(b). Neighbouring modules would be connected
so that their radiators fill a single volume.  Modules could be stacked
vertically, with hexagonal close packing, to make a wall.  Each wall would
then be followed by a tracking station, and this structure repeated as
often as necessary to provide the detector mass required.  Sixteen walls,
each of 61 modules, would provide a kiloton mass, as illustrated in
Fig~\ref{layout}\,(a).  Interleaving of toroidal magnets would allow the
muon momentum and charge to be determined.  The first tracking station
would act as a veto against charged particles entering the apparatus from
upstream, whilst the last station is separated by sufficient lever arm to
provide a measurement of the track angle after the last magnet.  The
required number of magnet and tracking stations would clearly be a matter
for detailed optimisation.

A more compact detector would be possible if a radiator of higher density
was available.  However, other possibilities that have been investigated
such as lead glass, whilst being suitably dense, also have a much higher
refractive index, so the volume imaged per detector is reduced (by Eq.\ 1).
Furthermore they lack the low chromatic error and large photon bandwidth of
$\rm C_6F_{14}$. Nevertheless, a suitable glass may still be found.

If the use of individual detectors for each module is abandoned, then the
module size could increase. The extreme case would be a large volume of
radiator limited by the dimensions of the experimental hall: for example, a
spherical mirror of radius 9\,m with a radiator length of 3\,m and an
array of detectors covering the upstream surface.  The photodetector
coverage would be a little lower, due to the packing of the tubes, and the
transparency of $\rm C_6F_{14}$ over such a long radiator length would need
to be studied. Also the large number of photons from a muon track would
give strong constraints on the mirror quality, to avoid a tail of poorly
reflected photons obscuring the tau signal.  The advantages are the
significantly reduced number of channels required, and the possibility of
using standard photodetectors.  One such module would have a radiator mass
of 240\,t, so could replace a series of four walls of modules in the
previous layout, as illustrated in Fig.~\ref{layout}\,(b).  The optimal
choice may lie somewhere between these two limits.

\section{Conclusions}

A novel concept for detection of tau neutrinos has been presented, through
their charged-current interaction in $\rm C_6F_{14}$ liquid to give a tau
lepton, that produces sufficient Cherenkov light for a ring image to be
formed. In about half of the events in a simple simulation a positive
identification of the tau can be achieved through the measurement of the
average Cherenkov angle of the detected photons.  The signature, for $\tau
\rightarrow \mu$ decays, is of a densely populated ring from the muon,
accompanied by an offset low intensity ring from the tau.  

Investigation of the pattern recognition issues, including the
effects of tracks from nuclear breakup in deep-inelastic interactions, will
await a more detailed simulation of the experiment.  Similarly, possible
background sources would need to be addressed, both technological (from
mirror imperfections, or backscattering from the silicon
of the HPD) and from physics (such as the production of delta rays, and
nuclear reinteraction). The purpose of this note is to gauge the interest
in the detector concept, before embarking on such a programme.

\section*{Acknowledgements}

It is a pleasure to thank Tom Ypsilantis for inspiration and advice: he
suggested the use of $\rm C_6F_{14}$ as radiator, and pointed me to Eq.\ 1.
Thanks also to Ioannis Papadopoulous and Pietro Antonioli, who provided the
simulated events used here, and to Christian Joram for the HPD information.

\end{document}